\documentclass[fleqn,11pt]{article}

\usepackage{amsmath,amsfonts,amssymb}
\usepackage{color}

\usepackage{geometry}
\newtheorem{theorem}{Theorem}
\newtheorem{lemma}[theorem]{Lemma}

\newtheorem{example}{Example}
\newtheorem{remark}{Remark}

\newcommand{\R}{\mathbb{R}}

\newcommand{\N}{\mathbb{N}}

\newcommand{\cqfd}
{%
\mbox{}%
\nolinebreak%
\hfill%
\rule{2mm}{2mm}%
\newline
\newline
}

\title{On stationary solutions to normal, coplanar discrete Boltzmann equation models.}
\author{Leif ARKERYD and Anne NOURI\\
\\Mathematical Sciences, 41296 G\"oteborg, Sweden,\\
arkeryd@chalmers.se\\
Aix-Marseille University, CNRS, Centrale Marseille, I2M UMR 7373, 13453 Marseille, France,\\
anne.nouri@univ-amu.fr}

\date{}

\begin{document}
\maketitle
\hspace{1cm}\\
{\bf Abstract} \hspace{1cm}\\
The paper proves existence of renormalized  solutions for a class of velocity-discrete  coplanar stationary Boltzmann equations with given indata. The proof is based on the construction of a sequence of approximations with $L^1$ compactness for the integrated
collision frequency and gain term. $L^1$ compactness of a sequence of approximations is obtained using the Kolmogorov-Riesz theorem and replaces the $L^1$ compactness of velocity averages in the continuous velocity case, not available when  the velocities are discrete.
\footnotetext[1] {2010 Mathematics Subject Classification; 60K35, 82C40, 82C99.} \\
\footnotetext[2] {Key words; stationary Boltzmann equation, discrete coplanar velocities, normal model, entropy.}
%
%
\section {Introduction.}
\label{generalization}
\setcounter{equation}{0}

The Boltzmann equation is the fundamental mathematical model in the kinetic theory of gases. Replacing its continuum of velocities with a discrete set of velocites is a  simplification  preserving the  essential features of free flow and quadratic collision term. Besides this fundamental aspect  they can  approximate the Boltzmann equation with any given accuracy \cite {BPS}, and are thereby useful for approximations and numerics. In the quantum realm they can also be more directly
connected to microscopic particles and quasiparticles models.\\
A discrete velocity model of a kinetic gas, is a system of partial differential equations having the form,
\begin{align*}
&\frac{\partial f_i}{\partial t}(t,z)+v_i\cdot \nabla _zf_i(t,z)= Q_i(f)(t,z),\quad t>0,\quad z\in \Omega ,\quad 1\leq i\leq p,
\end{align*}
where $f_i$, $1\leq i\leq p$, are phase space densities at time $t$, position $z$, velocity $v_i$, $\Omega \subset \R ^d$, and $v_i\in \R ^d$, $1\leq i\leq p$, are given discrete velocities. The collision operator $Q= (Q_i)_{1\leq i\leq p}$
with gain part $Q^+$, loss part $Q^-$, and collision frequency $\nu$, is given by
\begin{align*}
Q_i(f)&=\sum_{j,k,l=1}^p \Gamma_{ij}^{kl} (f_kf_l-f_if_j)\\
&= Q_i^+(f)-Q_i^-(f), \quad Q^-_i(f)= f_i\nu_i(f), \quad i=1,..., p.
\end{align*}
The collision coefficients satisfy 
\begin{align}\label{Gamma}
&\Gamma_{ij}^{kl}=\Gamma_{ji}^{kl}=\Gamma_{kl}^{ij}\geq 0.
\end{align}
If a  collision coefficient $\Gamma_{ij}^{kl}$ is non-zero, then the conservation laws for momentum and energy,
\begin{align}\label{conservations-velocities}
&v_i+v_j=v_k+v_l,\quad |v_i|^2+|v_j|^2=|v_k|^2+|v_l|^2,
\end{align}
are satisfied. The discrete velocity model (DVM) is called normal (see \cite{C}) if any solution of the equations
\begin{eqnarray*}
\Psi(v_i)+\Psi(v_j)=\Psi(v_k)+\Psi(v_l),
\end{eqnarray*}
where the indices $(i,j;k,l)$ take all possible values satisfying $\Gamma_{ij}^{kl}>0$, is given by 
\begin{align*}
&\Psi(v)=a+b\cdot v+c|v|^2,
\end{align*}
for some constants $a,c\in \R$ and $b\in\R^d$.
This paper studies stationary solutions to coplanar models, i.e. with $v_i\in \R ^2$, $1\leq i\leq p$, in a strictly convex bounded open subset $\Omega \subset \R ^2$, with  $C^1$ boundary $\partial \Omega $ and given indata. We consider the generic situation of normal coplanar velocities with
\begin{align}
&\text{no pair of  velocities  } v_i, v_j, 1\leq i,j\leq p, \text{  parallel},\label{generic-condition}\\
&\text {and additionally that for some direction } n_0\in \R ^2,\quad v_i\cdot n_0>0,\quad 1\leq i\leq p.\label{entropy2-main-theorem}
\end{align}
%
%
%
\begin{example}
\hspace*{1.in}\\
Let a model with velocities $v_i\in \R ^2$ satisfying \eqref{conservations-velocities}, $c_0>\mid v_i\mid  $, $1\leq i\leq p$, and $n_0\in \R ^2$ such that
\begin{align*}
&c_0n_0\notin \displaystyle{\cup _{1\leq i\neq j\leq p}}-v_j+\R (v_i-v_j).
\end{align*}
Then the model with velocities $v_i+c_0n_0$, $1\leq i\leq p$ satisfies \eqref{conservations-velocities}-\eqref{entropy2-main-theorem}.\\
Such a model based on the Broadwell model in the plane is
\begin{align*}
&(1,0)+(2,2),\quad (-1,0)+(2,2),\quad (0,1)+(2,2),\quad (0,-1)+(2,2).
\end{align*}
\end{example}
\begin{example}
\hspace*{1.in}\\
Discrete velocity models satisfying \eqref{conservations-velocities}-\eqref{entropy2-main-theorem} can also be constructed as follows. \\
Choose a direction $n_0\in \R ^2$. In the plane with origin $O$, denote by $P_+$ the half plane 
\begin{align*}
&P_+= \{ M\in \R ^2;\quad n_0\cdot \overrightarrow{OM}>0\} .
\end{align*}  
Choose $(A_i,A_j)\in P_+^2$, $A_i\neq A_j$, and $(A_l,A_m)\notin \{ (A_i,A_j), (A_j,A_i)\} $ diametrically opposed on the circle of diameter $[A_i,A_j]$. The quadrivector $(v_i,v_j,v_k,v_l)$ defined by
\begin{align*}
&v_i= \overrightarrow{OA_i},\quad v_j= \overrightarrow{OA_j},\quad v_l= \overrightarrow{OA_l},\quad v_m= \overrightarrow{OA_m},
\end{align*}
with a corresponding $\Gamma _{ij}^{kl}\neq 0$, satisfies \eqref{conservations-velocities}-\eqref{entropy2-main-theorem}. \\
Notice that some velocity can belong to different circles.
\end{example}
For stationary solutions to the Broadwell model, that does not belong to this class, see  \cite{AN2}, \cite{CIS}.\\
\hspace*{1.in}\\
Denote by $n(Z)$ the inward normal to $Z\in \partial \Omega $. Denote the $v_i$-ingoing (resp. $v_i$-outgoing) part of the boundary by 
\begin{align*}
&\partial \Omega _i^+= \{ Z\in \partial \Omega ;\quad  v_i\cdot n(Z)>0\}  ,\quad \quad (\text{resp.   }\partial \Omega _i^-= \{ Z\in \partial \Omega ;\quad  v_i\cdot n(Z)<0\}  ).
\end{align*}
Let
\begin{align*}
&s_i^+(z)= \inf \{ s>0;\hspace*{0.04in} z-sv_i\in \partial \Omega _i^+\} ,\quad  s_i^-(z)= \inf \{ s>0;\hspace*{0.04in} z+sv_i\in \partial \Omega _i^-\} ,\quad  z\in \Omega .
\end{align*}
Write 
\begin{align}\label{df-in-out-points}
z_i^+(z)= z-s_i^+(z)v_i \quad (\text{resp.   } z_i^-(z)= z+s_i^-(z)v_i)
\end{align}
for the ingoing (resp. outgoing) point on $\partial\Omega$ of the characteristics through $z$ in direction $v_i$. 
The boundary value problem
\begin{align}
&v_i\cdot \nabla f_i(z)= Q_i(f)(z),\quad z\in \Omega,\label{broad-general-a}\\
& f_i(z)= f_{bi}(z),\quad z\in \partial \Omega _i^+,\quad \quad 1\leq i\leq p,\label{broad-general-b}
\end{align}
is considered in $L^1$ in one of the following equivalent
forms (\cite{DPL});\\
the exponential multiplier form,
\begin{align}\label{exponential-form}
f_{i}(z)&= f_{bi}(z_i^+(z))e^{-\int_{0}^{s_i^+(z)} \nu_i(f)(z_i^+(z)+sv_i)ds}\nonumber\\
&+\int_{0}^{s_i^+(z)} Q_i^+(f)(z_i^+(z)+sv_i)e^{-\int_s^{s_i^+(z)}\nu_i(f)(z_i^+(z)+rv_i)dr}ds ,\quad \text{a.a.  }z\in \Omega ,\quad 1\leq i\leq p,
\end{align}
the mild form,
\begin{align}\label{mild-form}
&f_i(z)= f_{bi}(z_i^+(z))+\int_{0}^{s_i^+(z)} Q_i(f)(z_i^+(z)+sv_i)ds,\quad \text{a.a.  }z\in \Omega ,\quad 1\leq i\leq p,
\end{align}
the renormalized form,
\begin{align}\label{renormalized-form}
v_i\cdot \nabla \ln(1+f_i)(z)= \frac{Q_i(f)}{1+f_i}(z),\quad z\in \Omega ,\quad \quad f_i(z )= f_{bi}(z),\quad z\in \partial \Omega _i^+,\quad 1\leq i\leq p,
\end{align}
in  the sense of distributions.\\
Denote by $L^1_+(\Omega)$ the set of non-negative integrable functions on $\Omega$. Let
\begin{align}\label{df-entropy-dissipation}
&\sum _{(i,j,l,m)}\Gamma _{ij}^{lm}\int _\Omega (f_lf_m-f_if_j)\ln \frac{f_lf_m}{f_if_j}(z)dz
\end{align}
be the entropy dissipation of a distribution function $f$.
The main result of the present paper is 
%
%
\begin{theorem}\label{main-result}
\hspace*{1.in}\\
Consider a coplanar collision operator in the generic case of \eqref{generic-condition} additionally satisfying 
\eqref{entropy2-main-theorem},
and non-negative ingoing boundary values $f_{bi}$, $1\leq i\leq p$, with mass and entropy bounded,
\begin{align*}
&\int _{\partial \Omega _i^+}v_i\cdot n(z)\hspace*{0.02in}f_{bi}(1+\ln f_{bi})(z)d\sigma (z)<+\infty,\quad 1\leq i\leq p.
\end{align*} 
There exists a stationary renormalized solution in $\big( L^1_+(\Omega )\big) ^p$ to the boundary value problem \eqref{broad-general-a}-\eqref{broad-general-b} with finite mass, entropy and entropy-dissipation.
\end{theorem}
Most mathematical results for stationary discrete velocity models of the Boltzmann equation have been obtained in one space dimension. An overview is given in \cite{IP}. In two dimensions, special classes of solutions to the Broadwell model are given in \cite{CIS}, \cite{B},
and  \cite{Ily}. The Broadwell model is a four-velocity model, with $v_1+v_2=v_3+v_4=0$ and $v_1$, $v_2$ orthogonal. \cite{CIS} contains a detailed study of the stationary Broadwell equation in a rectangle with comparison to a
Carleman-like system, and a discussion of (in)compressibility aspects. A main result in \cite{CIS}
is the existence of continuous solutions to the two-dimensional stationary Broadwell model with continuous boundary data for a rectangle. The proof starts by solving the problem with a given gain term, and uses the compactness of the corresponding twice iterated  solution operator to conclude by Schaeffer's fixed point theorem. The paper \cite{AN2} studies that problem in an $L^1$-setting, with the proof  broadly within the frame of the present paper. In both those papers of ours, there is a priori control of mass and entropy dissipation.
Denoting by  $f_i(t,\cdot )$, $1\leq i\leq 4$, the density of the particles moving with velocity $v_i$ at time $t$, the proof in \cite{AN2} in an essential way uses the constancy of the sums  $f_1+f_2$ and $f_3+f_4$ along characteristics, which no longer holds in this paper. It is here replaced by a compactness property for the collision frequency and gain parts in the exponential form of the approximations employed. \\
The compactness is based on assumption \eqref{generic-condition} and the simultaneous presence of
space integrals in two velocity directions.\\
The proof starts from bounded  approximations with damping and convolution added, written in exponential multiplyer form, and solved by a fixed point argument.
Then the damping and convolutions are removed by taking limits using $L^1$-compactness of the integrated collision frequency and  gain term. The compactness  is proven by the Kolmogorov-Riesz theorem (see \cite{K}, \cite{R}). The limit of the remaining approximations is obtained by using again the Kolmogorov-Riesz theorem.
%
%
%
%
\section{Approximations.}
\label{approximations}
\setcounter{equation}{0}
\setcounter{theorem}{0}
The construction of the primary approximated boundary value problem with damping and convolutions is similar to the Broadwell case \cite{AN2} and given in the following lemma. Denote by $a\wedge b$ the minimum  of two real numbers $a$ and $b$. Take  $\alpha >0$ and set
\begin{align}\label{df-c-alpha}
&c_\alpha = \frac{1}{\alpha }\sum _{i=1}^p\int _{\partial\Omega _i^+}(n(z)\cdot v_i)f_{bi}(z)d\sigma (z),\quad K_\alpha=\{f\in \big( L_+^1(\Omega)\big) ^p;\sum _{i=1}^p\int_\Omega f_i(z)dz\leq c_\alpha\} .
\end{align}
Let $\mu_\alpha$ be a smooth mollifier in $\R^2$
with support in the ball centered at the origin of radius $\alpha $. Outside the boundary the function to be convolved with $\mu_\alpha$ is continued in the normal direction by its boundary value. Let $\tilde{\mu }_k$ be a smooth mollifier on $\partial \Omega $. Denote by 
\begin{align*}
&f^k_{bi}= \Big( f_{bi}(\cdot )\wedge \frac{k}{2}\Big) \ast \tilde{\mu }_k ,\quad 1\leq i\leq p.
\end{align*}
\begin{lemma}\label{first-approximations}
 There is a solution $F\in (L^1_+(\Omega))^p$ to 
\begin{align}
&\alpha F_i+v_i\cdot \nabla F_i= \sum _{j,l,m=1}^p\Gamma _{ij}^{lm}\Big( \frac{F_l}{1+\frac{F_l}{k}}\frac{F_m\ast \mu_\alpha }{1+\frac{F_m\ast \mu_\alpha }{k}}-\frac{F_i}{1+\frac{F_i}{k}}\frac{F_j\ast \mu_\alpha }{1+\frac{F_j\ast \mu_\alpha }{k}}\Big)\hspace*{0.03in},\label{df-F-final-1}\\
&F_i(z_i^+(z))= f^k_{bi}(z_i^+(z)),\quad 1\leq i\leq p.\label{df-F-final-2}
\end{align}
\end{lemma}
\underline{Proof of Lemma \ref{first-approximations}.}
Let $\mathcal{T}$ be  the map defined on $K_\alpha $ by $\mathcal{T}(f)=F$, where $F= (F_i)_{1\leq i\leq p}$ is the solution of
\begin{align}
&\alpha F_i+v_i\cdot \nabla F_i= \sum _{j,l,m=1}^p\Gamma _{ij}^{lm}\Big( \frac{F_l}{1+\frac{F_l}{k}}\frac{f_m\ast \mu_\alpha }{1+\frac{f_m\ast \mu_\alpha }{k}}-\frac{F_i}{1+\frac{F_i}{k}}\frac{f_j\ast \mu_\alpha }{1+\frac{f_j\ast \mu_\alpha }{k}}\Big)\hspace*{0.03in},\label{df-F-1}\\
&F_i(z_i^+(z))= f^k_{bi}(z_i^+(z)).\label{df-F-2}
\end{align}
\\
$F= \mathcal{T}(f)$ can be obtained as the limit in $(L_+^1(\Omega))^p$ of the sequence $(F^q)_{q\in \N }$ defined by $F^0= 0$ and
\begin{align}
&\alpha F^{q+1}_i+v_i\cdot \nabla F^{q+1}_i= \sum _{j,l,m=1}^p\Gamma _{ij}^{lm}\Big( \frac{F^q_l}{1+\frac{F^q_l}{k}}\frac{f_m\ast \mu_\alpha }{1+\frac{f_m\ast \mu_\alpha }{k}}-\frac{F^{q+1}_i}{1+\frac{F^{q}_i}{k}}\frac{f_j\ast \mu_\alpha }{1+\frac{f_j\ast \mu_\alpha }{k}}\Big)\hspace*{0.03in},\label{df-F1}\\
&F^{q+1}_i(z_i^+(z))= f^k_{bi}(z_i^+(z))\hspace*{0.03in},\hspace*{3.2in}q\in \N.
\end{align}
$F^{q+1}$ can be written in the following exponential form,
\begin{align}\label{exp-form-Fq}
&F_i^{q+1}(z)= f^k_{bi}(z_i^+(z))e^{-\alpha s_i^+(z)-\sum _{j,l,m}\Gamma _{ij}^{lm}\int _{-s_i^+(z)}^0\frac{f_j\ast \mu_\alpha }{(1+\frac{F^{q}_i}{k})(1+\frac{f_j\ast \mu_\alpha }{k})}(z+sv_i)ds}\nonumber \\
&+\sum _{j,l,m=1}^p\Gamma _{ij}^{lm}\int _{-s_i^+(z)}^0\frac{F^q_l}{1+\frac{F^q_l}{k}}\frac{f_m\ast \mu_\alpha }{1+\frac{f_m\ast \mu_\alpha }{k}}(z+sv_i)e^{\alpha s-\sum _{j,l,m}\Gamma _{ij}^{lm}\int _{s}^0\frac{f_j\ast \mu_\alpha }{(1+\frac{F^{q}_i}{k})(1+\frac{f_j\ast \mu_\alpha }{k})}(z+rv_i)dr}ds,\nonumber \\
&\hspace*{4.5in}1\leq i\leq p.
\end{align}
 \\
The sequence $(F^q)_{q\in \N }$ is monotone. Indeed, 
\begin{align*}
&F_i^0\leq F_i^1,\quad 1\leq i\leq n,
\end{align*} 
by the exponential form of $F_i^1$. If $F_i^q\leq F_i^{q+1}$, $1\leq i\leq p$, then it follows from the exponential form
that $F_i^{q+1}\leq F_i^{q+2}$.
Moreover,
\begin{align*}
&\alpha \sum_{i=1}^pF^{q+1}_i+\sum_{i=1}^pv_i\cdot \nabla F^{q+1}_i= \sum _{i,j,l,m=1}^p\Gamma _{ij}^{lm} \frac{(F^q_l-F^{q+1} _{l})}{1+\frac{F^q_l}{k}}\frac{f_m\ast \mu_\alpha }{1+\frac{f_m\ast \mu_\alpha }{k}}
\leq 0,
\end{align*}
so that
\begin{align}\label{total-mass-Fn-i}
&\sum _{i=1}^p\int _\Omega F_i^{q+1}(z)dz\leq c_\alpha .
\end{align}
By the monotone convergence theorem, $(F^q)_{q\in \N }$ converges in $L^1(\Omega)$ to a solution $F$ of \eqref{df-F-1}-\eqref{df-F-2}. The solution of \eqref{df-F-1}-\eqref{df-F-2} is unique in the set of non-negative functions. Indeed, let $G=(G_i)_{1\leq i\leq p}$ be a non-negative solution of \eqref{df-F-1}-\eqref{df-F-2}. It follows by induction that
\begin{align}\label{recurrence1}
&\forall q \in \N , \quad F_i^q \leq G_i,\quad 1\leq i\leq p.
\end{align}
Indeed, \eqref{recurrence1} holds for $q= 0$, since $G_i\geq 0$, $1\leq i\leq p$. Assume \eqref{recurrence1} holds for $q$. Using the exponential form of
$F_i^{q+1}$ implies $F_i^{q+1}\leq G_i$.
Consequently,
\begin{align}\label{comparaison-F-G}
&F_i\leq G_i,\quad 1\leq i\leq p.
\end{align}
Moreover, subtracting the partial differential equations satisfied by $G_i$ from the partial differential equations satisfied by $F_i$, $1\leq i\leq p$, and integrating the resulting equation on $\Omega$, it results
\begin{align}\label{total-mass-F-i}
\alpha \sum _{i=1}^p\int_\Omega (G_i-F_i)(z)dz
+\sum _{i=1}^p\int_{\partial \Omega _i^-} |n(z)\cdot v_i|(G_i-F_i)(z)d\sigma (z)= 0.
\end{align}
It results from \eqref{comparaison-F-G}-\eqref{total-mass-F-i} that $G= F$.\\
The map $\mathcal{T}$ is continuous in the $L^1$-norm topology (cf \cite{AN} pages 124-5). Namely, let a sequence $(f^q)_{q\in \N }$ in $K_\alpha$ converge in $(L^1(\Omega))^p$ to $f\in K_\alpha$. Set $F^q= \mathcal{T}(f^q)$. Because of the uniqueness of the solution to \eqref{df-F-1}-\eqref{df-F-2}, it is enough to prove that there is a subsequence of  $(F^q)$ converging to $F= \mathcal{T}(f)$. Now there is a subsequence of $(f^q)$, still denoted $(f^q)$, such that decreasingly (resp. increasingly) $(G^q)= (\sup _{r \geq q}f^r)$ (resp. $(g^q)= (\inf _{r\geq q}f^r)$)  converges to $f$ in $L^1$. Let $(S^q)$ (resp. $(s^q)$) be the sequence of solutions to
\begin{align}
&\alpha S^q_i+v_i\cdot \nabla S^q_i= \sum _{j,l,m=1}^p\Gamma _{ij}^{lm}\Big( \frac{S^q_l}{1+\frac{S^q_l}{k}}\frac{G^q_m\ast \mu_\alpha }{1+\frac{G^q_m\ast \mu_\alpha }{k}}-\frac{S^q_i}{1+\frac{S^q_i}{k}}\frac{g^q_j\ast \mu_\alpha }{1+\frac{g^q_j\ast \mu_\alpha }{k}}\Big)\hspace*{0.03in},\label{df-F1}\\
&S^q_i(z_i^+(z))= f^k_{bi}(z_i^+(z)),
\end{align}
\begin{align}
&\alpha s^q_i+v_i\cdot \nabla _xs^q_i= \sum _{j,l,m=1}^p\Gamma _{ij}^{lm}\Big( \frac{s^q_l}{1+\frac{s^q_l}{k}}\frac{g^q_m \ast \mu_\alpha }{1+\frac{g^q_m\ast \mu_\alpha }{k}}-\frac{s^q_i}{1+\frac{s^q_i}{k}}\frac{G^q_j\ast \mu_\alpha }{1+\frac{G^q_j\ast \mu_\alpha }{k}}\Big)\hspace*{0.03in},\label{df-F1}\\
&s^q_i(z_i^+(z))= f^k_{bi}(z_i^+(z))\hspace*{0.03in}.
\end{align}
$(S^q)$ is a non-increasing sequence, since that holds for the successive iterates defining the sequence. Then $(S^q)$ decreasingly converges in $L^1$ to some $S$. Similarly $(s^q)$ increasingly converges in $L^1$ to some $s$. The limits $S$ and $s$ satisfy \eqref{df-F-1}-\eqref{df-F-2}. It follows by uniqueness that $s= F= S$, hence that $(F^q)$ converges in $L^1$ to $F$.\\
The map $\mathcal{T}$ is also compact in the $L^1$-norm topology. Indeed, let $(f^q)_{q \in \N }$ be a sequence in $K_\alpha$ and $(F^q)_{q \in \N}= (\mathcal{T}(f^q))_{q \in \N }$.
The boundedness by $k^2$ of the terms in the collision operator,
 induces uniform $L^1$ equi-continuity of $(F^q_i)_{q\in \N }$
 with respect to the $v_i$-direction, as follows from the mild form of the equations.  
For the uniform $L^1$ equi-continuity with respect to the $v_j$-direction, $j\neq i$, 
consider for each $q$ and with $f:=f^q$ the sequence $(G^{q,r})_{r\in \N }$ defined by $G^{q,0}=0$  and for $r\in \N ^*$
 \begin{align}
&\alpha G^{q,r}_i+v_i\cdot \nabla G^{q,r}_i= \sum _{j,l,m=1}^p\Gamma _{ij}^{lm}\Big( \frac{G^{q,r}_l}{1+\frac{G^{q,r}_l}{k}}\frac{f_m\ast {\mu}_\alpha }{1+\frac{f_m\ast {\mu}_\alpha }{k}}-\frac{G^{q,r}_i}{1+\frac{G^{q,r-1}_i}{k}}\frac{f_j\ast {\mu}_\alpha }{1+\frac{f_j\ast {\mu}_\alpha }{k}}\Big)\hspace*{0.03in},\label{df-G-1}\\
&G^{q,r}_i(z_i^+(z))= f^k_{bi}(z_i^+(z))),\quad 1\leq i\leq p.\label{df-G-2}
\end{align}
The existence of a unique solution for each $r$ follows as for the problem \eqref{df-F-1}-\eqref{df-F-2}. By induction on $r$, prove that $(G^{q,r})_{q\in \N }$ is uniformly equicontinuous in the $v_j$-direction. It holds for $r= 0$. Assume it holds for $r-1\in \N ^*$ and prove it for $r$. Writing $G^{q,r}(z)$ in exponential form and using the uniform equicontinuity in the $v_j$-direction of $(G^{q,r-1})_{q\in \N }$ and the compactness of $(f^q\ast \mu _\alpha )$, it comes back to prove the uniform equicontinuity in the $v_j$-direction of 
\begin{align*}
&z\rightarrow \int _0^{s_i(z_i^-(z))} \frac{G_l^{q,r}}{1+\frac{G_l^{q,r}}{k}}(z_i^-(z)+sv_i)ds.
 \end{align*}
First,
\begin{align*}
&z_i^-(z+hv_j)= z_i^-(z)+av_i+bv_l,\quad \text{with}\quad \lim _{h\rightarrow 0}a(h)= \lim _{h\rightarrow 0}b(h)= 0,
\end{align*}
uniformly with respect to $z\in \Omega $. Consequently,
\begin{align}
&\int \mid \int _0^{s_i(z_i^-(z))} \frac{G_l^{q,r}}{1+\frac{G_l^{q,r}}{k}} (z_i^-(z+hv_j)+sv_i)-\frac{G_l^{q,r}}{1+\frac{G_l^{q,r}}{k}} (z_i^-(z)+sv_i)ds\mid dz\nonumber \\
&\leq \int \int _0^{s_i(z_i^-(z))} \mid \frac{G_l^{q,r}}{1+\frac{G_l^{q,r}}{k}} (z_i^-(z)+(a+s)v_i+bv_l)-\frac{G_l^{q,r}}{1+\frac{G_l^{q,r}}{k}} (z_i^-(z)+(a+s)v_i)\mid dsdz\label{compactnessT-1}\\
&+\int \mid \int _0^{s_i(z_i^-(z))} \frac{G_l^{q,r}}{1+\frac{G_l^{q,r}}{k}} (z_i^-(z)+(a+s)v_i)-\frac{G_l^{q,r}}{1+\frac{G_l^{q,r}}{k}} (z_i^-(z)+sv_i)ds\mid dz.\label{compactnessT-2}
\end{align}
The limit when $h\rightarrow 0$ of \eqref{compactnessT-1} is zero, by the uniform $L^1$ equicontinuity of $(G^{q,r}_l)_{q \in \N }$ with respect to the $v_l$-direction.
With the change of variables $s\rightarrow a+s$ in its first integral, \eqref{compactnessT-2} equals
\begin{align}
&\int \mid \int _{s_i(z_i^-(z))}^{s_i(z_i^-(z))+a} \frac{G_l^{q,r}}{1+\frac{G_l^{q,r}}{k}} (z_i^-(z)+sv_i)ds-\int _0^a\frac{G_l^{q,r}}{1+\frac{G_l^{q,r}}{k}} (z_i^-(z)+sv_i)ds\mid dz,
\end{align}
which tends to zero when $h$ tends to zero since both integrands are bounded by $k$. This proves the $L^1$ compactness of $(G^{q,r}_i)_{q \in \N}$. For $q$ fixed, the sequence  $(G^{q,r}_i)_{r \in \N}$ is increasing, and its limit satisfies \eqref{df-F-1}-\eqref{df-F-2} with $f=f^q$, so  the limit equals $F^q_i$. Take a subsequence of $q$ still denoted by $q$, with
$(f^q\ast \mu_\alpha )$ convergent in $L^1$ to some $f^\infty$ when $q\rightarrow \infty $, and a further subsequence so that $(G^{q,1})$ converges to some $F^{\infty ,1}$ in $L^1$. Continue by diagonalization to convergence of $(G^{q,r})_q$ to $F^{\infty ,r}$ for all $r\in \N $. The limits satisfy \eqref{df-G-1}-\eqref{df-G-2} with $f\ast \mu_\alpha$ replaced with $f^\infty$, and $G^{q,r}$ with $F^{\infty,r}$ 
giving an increasing sequence,  with limit satisfying \eqref{df-F-1}-\eqref{df-F-2}, where $f\ast\mu_\alpha$  is replaced with $f^\infty$.
So given a sequence in $K_\alpha$, there is a subsequence with converging image under $\mathcal{T}$.
%
The compactness of $\mathcal{T}$ is thus proved.\\
Hence by the Schauder fixed point theorem, there is a fixed point for $\mathcal{T}$, i.e. a solution $F$ to \eqref{df-F-final-1}-\eqref{df-F-final-2}.
\cqfd
%
%
%
%
\section{Removal of the damping and convolutions.}
\setcounter{equation}{0}
\setcounter{theorem}{0}
Let $k>1$ be fixed. Denote by $F^{\alpha ,k}$ the solution to \eqref{df-F-final-1}-\eqref{df-F-final-2} obtained in the previous section. Each component of $F^{\alpha ,k}$ being bounded by a multiple of $k^2$, $(F^{\alpha ,k})_{\alpha \in ]0,1[}$ is weakly compact in $(L^1(\Omega))^p$. Denote by $F^k$ the limit  for the weak topology in $(L^1(\Omega))^p$ of a converging subsequence when $\alpha\rightarrow 0$. Let us prove that for a subsequence, the convergence is strong in $(L^1(\Omega))^p$.
%
%
\begin{lemma}\label{cv-F-alpha}
{There is a sequence $(\alpha _q)_{q\in \N}$ tending to zero when $q\rightarrow +\infty $,  such that $(F^{\alpha _q,k})_{q\in \N }$ } strongly converges to $F^k$ in $(L^1(\Omega))^p$ when $q\rightarrow +\infty $.
\end{lemma}
\underline{Proof of Lemma \ref{cv-F-alpha}}\\
Consider the approximation scheme $(f^{\alpha,\rho })_{\rho \in \N }$ of $F^{\alpha ,k}$,

\begin{align}
&f^{\alpha, 0}_i=0,\\
&\alpha f_i^{\alpha,\rho+1} +v_i\cdot \nabla f_i^{\alpha, \rho+1}= \sum _{j,l,m=1}^p\Gamma _{ij}^{lm}\Big( \frac{F^{\alpha ,k}_l}{1+\frac{F^{\alpha ,k}_l}{k}}\frac{F^{\alpha ,k}_m\ast \mu_\alpha }{1+\frac{F^{\alpha ,k}_m\ast \mu_\alpha }{k}}-\frac{f^{\alpha,\rho+1}_i}{1+\frac{f^{\alpha,\rho+1}_i}{k}}\frac{f^{\alpha,\rho }_j\ast \mu_\alpha }{1+\frac{f^{\alpha,\rho }_j\ast \mu_\alpha }{k}}\Big)\hspace*{0.03in},\label{df-F1}\\
&f^{\alpha,\rho+1}_i(z^+_i(z))= f^k_{bi}(z^+_i(z)),\quad 1\leq i\leq p, \quad \rho \in \N.
\end{align}
$f^{\alpha ,1}$ is obviously given in terms of $F^{\alpha ,k}$. It follows from the exponential form that  
\begin{align*}
&F_i^{\alpha ,k}\leq f_i^{\alpha,1},\quad \alpha \in ]0,1[.
\end{align*}
Denote by $\mathcal{S}$ the map from $\R ^p\times \R^p$ mapping $(X,Z)$ into $W= \mathcal{S}(X,Z)\in \R ^p$ solution to 
\begin{align*}
&\alpha W_i+v_i\cdot \nabla W_i= \sum _{j,l,m=1}^p\Gamma _{ij}^{lm}\Big( \frac{F^{\alpha ,k}_l}{1+\frac{F^{\alpha ,k}_l}{k}}\frac{F^{\alpha ,k}_m\ast \mu_\alpha }{1+\frac{F^{\alpha ,k}_m\ast \mu_\alpha }{k}}-\frac{W_i}{1+\frac{X_i}{k}}\frac{Z_j\ast \mu_\alpha }{1+\frac{Z_j\ast \mu_\alpha }{k}}\Big)\hspace*{0.03in},\\
&W_i(z^+_i(z))= f^k_{bi}(z^+_i(z)),\quad 1\leq i\leq p.
\end{align*} 
Denote by
\begin{align*}
&f^{\alpha,1,0}= \mathcal{S}(0,f^{\alpha,1}),\quad f^{\alpha,1,r}= \mathcal{S}(f^{\alpha,1,r-1},f^{\alpha,1}),\\
&F^{\alpha, k,0}= \mathcal{S}(0,F^{\alpha ,k}),\quad F^{\alpha ,k, r}= \mathcal{S}(F^{\alpha ,k, r-1}, F^{\alpha ,k}),\quad r\in \N ^*.
\end{align*}
First, 
\begin{align*}
&f_i^{\alpha,1,0}\leq F_i^{\alpha,k,0}.
\end{align*}
Then the sequence  $(f_i^{\alpha,1,r})_{r\in \N }$ (resp. $(F_i^{\alpha,k,r})_{r\in \N }$) is increasing with limit $f_i^{\alpha,2}$ (resp. $F_i^{\alpha ,k}$).
It follows from $f_i^{\alpha,1,r}\leq F_i^{\alpha,k,r}$, $r\in \N $, that 
\begin{equation}\label{comparison}
f_i^{\alpha,2}\leq F_i^{\alpha ,k},\quad 1\leq i\leq p.
\end{equation}
Let
\begin{align*}
&f^{\alpha,2,0}:= \mathcal{S}(0,f^{\alpha,2}),\quad f^{\alpha,2,r}:= \mathcal{S}(f^{\alpha,r-1},f^{\alpha ,2}),\quad r\in \N ^*.
\end{align*}  
It follows from \eqref{comparison} that 
\begin{align*}
&f_i^{\alpha,2,0}\geq F_i^{\alpha , k,0},\quad 1\leq i\leq p.
\end{align*} 
The sequence $(f_i^{\alpha,2,r})_{r\in \N }$ is also increasing with limit $f_i^{\alpha,3}$ and with $f_i^{\alpha,2,r}\geq F_i^{\alpha ,k,r}$.  Hence 
\begin{align*}
&f_i^{\alpha,3}\geq F_i^{\alpha ,k}.
\end {align*}
From here by induction on $\rho $, it holds that
\begin{align}\label{order-f-alpha-l}
&f_i^{\alpha ,2\rho }\leq f_i^{\alpha ,2\rho +2}\leq F_i^{\alpha ,k}\leq f_i^{\alpha ,2\rho +3}\leq f_i^{\alpha ,2\rho+1},
\quad \alpha \in ]0,1[ ,\quad \rho\in \N .
\end{align}
By induction on $r$, for each $r$ the sequence 
$(f^{\alpha,1,r})_{\alpha \in ]0,1[}$ is translationally equicontinuous in $\alpha$. The limit sequence $(f^{\alpha,2})_{\alpha \in ]0,1[}$ is also translationally equicontinuous. This is so, since given $\epsilon>0$, 
$r$ and then $h_0$ can be taken so that 
\begin{align*}
\int (f^{\alpha,2}-f^{\alpha,1,r})(z)dz<\epsilon \quad  \text{and }\quad \int |f^{\alpha,1,r}(z+h)-f^{\alpha,1,r}(z)|dz<\epsilon , \quad |h|<h_0.
\end{align*}
It can analogously be proven that for each $\rho \in \N $, $(f^{\alpha ,\rho })_{\alpha \in ]0,1[}$ is translationally equicontinuous in $\alpha $. Let $(\alpha _q)_{q\in \N }$ be a sequence tending to zero. Take a subsequence in $(\alpha _q)_{q\in \N}$, still denoted by $(\alpha _q)_{q\in \N}$, such that $(f^{\alpha _q,2})_{q\in \N }$  converges in $L^1$ to some $f^{0,2}$ when $q \rightarrow +\infty $.\\
\hspace*{1.in}\\
Continuing by induction gives a sequence $(f^{0,\rho })_{\rho \in \N}$ 
satisfying  
\begin{align}
&f_i^{0 ,2\rho }\leq f_i^{0 ,2\rho +2}\leq F_i^k \leq f_i^{0 ,2\rho +3}\leq f_i^{0 ,2\rho +1},
\quad \rho \in \N ,
\end{align}
\begin{align*}
&v_i\cdot \nabla f^{0,\rho +1}_i=  G_i-
\sum _{j,l,m=1}^p\Gamma _{ij}^{lm}\frac{f^{0,\rho +1}_i}{1+\frac{f^{0,\rho +1}_i}{k}}\frac{f^{0,\rho }_j }{1+\frac{f^{0\rho }_j}{k}},
\\
&f^{0,\rho +1} _i(z^+_i(z))= f^k_{bi}(z^+_i(z)).
\end{align*}
Here,  $G_i^k$ is the weak $L^1$ limit when $\alpha \rightarrow 0$ of the gain term
\begin{align*}
&\sum _{j,l,m=1}^p\Gamma _{ij}^{lm} \frac{F^{\alpha ,k}_l}{1+\frac{F^{\alpha ,k}_l}{k}}\frac{F^{\alpha ,k}_m\ast \mu_\alpha }{1+\frac{F^{\alpha ,k}_m\ast \mu_\alpha }{k}}\hspace*{0.04in}.
\end{align*}
In particular, $(f^{0,2\rho }_i)_{\rho \in \N }$ 
 (resp. $(f^{0,2\rho +1}_i)_{\rho  \in \N }$) non decreasingly (resp. non increasingly) converges in $L^1$ to some $g_i$ (resp. $h_i$) when $\rho \rightarrow +\infty $. The limits satisfy
\begin{align}
&0\leq g_i\leq F^k_i\leq h_i,\nonumber \\
&v_i\cdot \nabla h_i=  G_i-\sum _{j,l,m=1}^p\Gamma _{ij}^{lm}\frac{h_i}{1+\frac{h_i}{k}}\frac{g_j }{1+\frac{g_j}{k}},\label{eq-hi}\\
& v_i\cdot \nabla g_i=  G_i-\sum _{j,l,m=1}^p\Gamma _{ij}^{lm}\frac{g_i}{1+\frac{g_i}{k}}\frac{h_j }{1+\frac{h_j}{k}},\label{eq-gi}\\
&(h_i-g_i)(z_i^+(z))= 0.\nonumber 
\end{align}
Integrating and summing gives that 
\begin{align*}
&\sum_{i=1}^p\int_{\partial \Omega _i^-}\mid v_i\cdot n(Z)\mid (h_i-g_i)(Z)d\sigma (Z)= 0,
\end{align*} 
i.e. that $g_i= h_i$ also on $\partial\Omega _i^-$. 
Integrating the equation satisfied by $h_i-g_i$ over the part of $\Omega $ on one side of a line orthogonal to $n_0$, summing over $i$ and using \eqref{entropy2-main-theorem} implies that $g=h $ on that line, hence in all  of $\Omega$, and is equal to $F_i^k$.
$(F^{\alpha _q, k})_{q\in \N }$ 
converges to $F^k$ in $(L^1(\Omega))^p$ when $q\rightarrow +\infty $. Indeed, given $\eta >0$, choose $\rho _0$ big enough so that 
\begin{align*}
&\parallel f^{0,2\rho _0+1}_i-f^{0,2\rho _0}_i\parallel _{L^1}<\eta \quad \text{and}\quad \parallel f^{0,2\rho _0}_i-F^k_i\parallel _{L^1}<\eta ,\quad 1\leq i\leq p,
\end{align*}
 then $q_0$ big enough, so that
\begin{align*}
&\parallel f^{\alpha _q, 2\rho _0+1}_i-f^{0,2\rho _0+1}_i\parallel _{L^1}\leq \eta \quad \text{and}\quad \parallel f^{\alpha _q,2\rho  _0}_i-f^{0,2\rho _0}_i\parallel _{L^1}\leq \eta ,\quad q\geq q_0.
\end{align*}
Then split $\parallel F^{\alpha _q, k} _i-F^k_i\parallel _{L^1}$ as follows,
\begin{align*}
&\parallel F^{\alpha _q, k}_i-F^k_i\parallel_ {L^1}\\
&\leq \parallel F^{\alpha _q, k}_i-f^{\alpha ,2\rho _0}_i\parallel _{L^1}+\parallel f^{\alpha ,2\rho _0}_i-f^{0,2\rho _0}_i\parallel _{L^1}+\parallel f^{0,2\rho _0}_i-F^k_i\parallel _{L^1}\nonumber \\
& \leq \parallel f^{\alpha , 2\rho _0+1}_i-f^{\alpha ,2\rho _0}_i\parallel _{L^1}+2\eta \quad \text{by}\quad \eqref{order-f-alpha-l}\nonumber \\
& \leq \parallel f^{\alpha , 2\rho _0+1}_i-f^{0,2\rho _0+1}_i\parallel _{L^1}+\parallel f^{0,2\rho _0+1}_i-f^{0,2\rho _0}_i\parallel _{L^1}+\parallel f^{0,2\rho _0}_i-f_i^{\alpha ,2\rho _0}\parallel _{L^1}+2\eta \nonumber \\
&\leq 5\eta ,\quad q\geq q_0.
\end{align*}
\cqfd 
%
%
\begin{lemma}\label{existence-k-approximations}
For any $k\in \N ^*$, $F^k$ is a nonnegative continuous solution to
\begin{align}
&v_i\cdot \nabla F_i^k=Q_i^{+k}-F_i^k\nu_i^k
\hspace*{0.03in},\label{Fk-i}\\
&F^k_i(z^+_i(z))= f^k_{bi}(z^+_i(z)),\quad 1\leq i\leq p,\label{bcFk-i}
\end{align}
where
\begin{align*}
&Q^{+k}_i= \sum _{j,l,m=1}^p\Gamma _{ij}^{lm}\frac{F^k_l}{1+\frac{F^k_l}{k}}\frac{F^k_m}{1+\frac{F^k_m}{k}},\quad \nu _i^k= \sum _{j,l,m=1}^p\Gamma _{ij}^{lm}\frac{F^k_j}{(1+\frac{F^k_i}{k})(1+\frac{F^k_j}{k})}\hspace*{0.04in}.
\end{align*}
\end{lemma}
%
%
\underline{Proof of Lemma \ref{existence-k-approximations}.}\\
Passing to the limit when $q\rightarrow +\infty $ in \eqref{df-F-final-1}-\eqref{df-F-final-2} written for $F^{\alpha _q, k}$, implies that $F^k$ is a solution in $(L^1_+(\Omega ))^p$ to \eqref{Fk-i}-\eqref{bcFk-i}. It remains to prove its continuity. Using twice its exponential form and the continuity of $f^k_{b}$, it comes back to prove the continuity of
\begin{align}\label{pf-lemma-3.1-a}
&\int _0^{s_i^+(z)}\int _0^{s_j^+(z_i^+(z)+sv_i)}G^k(z_j^+(z_i^+(z)+sv_i)+\sigma v_j)d\sigma ds,\quad i\neq j,
\end{align}
for given measurable bounded functions $G^k$. The mapping
\begin{align}\label{change-variables}
&(s, \sigma )\in [0, s_i^+(z)] \times [0, s_j^+(z_i^+(z)+sv_i)] \rightarrow Z= z_j^+(z_i^+(z)+sv_i)+\sigma v_j,
\end{align}
is a change of variables. Indeed, the strict convexity of $\Omega $ and the $C^1$ regularity of $\partial \Omega $ imply that $z\rightarrow z_i^+(z)$ is well-defined and $C^1$ for any $i\in \{ 1,\cdot \cdot \cdot , p\} $. Hence the map $(s,\sigma )\rightarrow Z$ is one to one and $C^1$. Its Jacobian equals one since $Z= Z_iv_i+Z_jv_j$, with $Z_i= s-s_i^+(z) $ linear in $s$ and independent of $\sigma$, and $Z_j= \sigma -s_j^+\big( z+(s-s_i^+(z))v_i\big)$ linear in $\sigma$.\\
Using this change of variable leads to the continuity of the map defined in \eqref{pf-lemma-3.1-a}.                     \cqfd
%
 %
\begin{lemma}\label{mass-entropy-dissipation-approximations}
Solutions $(F^k)_{k\in \N^*}$ to \eqref{Fk-i}-\eqref{bcFk-i} have mass and entropy dissipation bounded from above uniformly with respect to $k$.
\end{lemma}
\underline{Proof of Lemma \ref{mass-entropy-dissipation-approximations}.}\\
Choose an orthonormal basis $(e_x,e_y)$ of $\R^2$ so that
neither the $x$-direction nor the $y$-direction is parallel to any of $v_1$, ..., $v_p$.
Observe that integrating \eqref{Fk-i}-\eqref{bcFk-i} over $\Omega$ and summing over $i$, shows that outflow of mass equals inflow.
 We shall  first obtain uniformly in $k$, an upper bound for the energy 
 \begin{align*} 
 &\sum _{i=1}^pv_i^2\int _\Omega F^k_i(z)dz.
 \end{align*}
Recalling that the genericity condition \eqref{generic-condition} implies that all velocities are different from zero, the energy bound
 implies an upper estimate for the mass. Write $v_i= \xi_i e_x+\zeta_i e_y$. Multiply the equation for $F^k_i$ with $\xi_i$ and integrate over $\Omega_a=\Omega\cap
 \{(x,y); x\leq a\}$. Set
 \begin{align*}
 &S_a=\Omega\cap\{(x,y);x=a\} \quad \text{and}\quad \partial\Omega_a=\partial
 \Omega\cap \bar{\Omega}_a.
 \end{align*}   
 From \eqref{Fk-i}-\eqref{bcFk-i} follows
 \begin{align}\label{lemma3.2-a}
 &\sum_{i=1}^p\xi_i^2\int_{S_a}F^k_i(a,y)dy= \sum_{i=1}^p\xi_i\int_{\partial\Omega_a}(v_i\cdot n(Z))F^k_i(Z)d\sigma (Z).
 \end{align}
For any $(x,y)\in\Omega$ let the line-segment through  $(x,y)$ in the $x$-direction (resp. $y$-direction) intersect the boundary $\partial \Omega$ at $x^-(y)<x^+(y)$ (resp. $y^-(x)<y^+(x)$). Denote by
\begin{align}\label{df-x0-x0+}
&x_0^-:= \min_{ (x,y)\in \Omega }\{x^-(y)\} , \quad x_0^+:= \max_{(x,y)\in \Omega }\{x^+(y)\} .
\end{align}
 Integrating \eqref{lemma3.2-a} on  $a\in [x_0 ^-,x_0^+]$ gives uniformly in $k$,
 \begin{align*}
& \sum_{i=1}^p\xi_i^2\int_{\Omega}F_i^k(z)dz= \sum_{i=1}^p\xi_i\int_{x_0^-}^{x_0^+}\Big( \int_{\partial\Omega_a} (v_i\cdot n(Z))F^k_i(Z)d\sigma (Z)\Big) da\leq c_b,
 \end{align*}
where $c_b$ only depends on the given inflow. Analogously $\sum_{i=1}^p\zeta _i^2\int_ \Omega F^k_i(z)dz\leq c_b$. The boundedness of energy and with it mass, follows.\\
The entropy dissipation estimate is proved  as follows. Denote by $D^k$ the entropy dissipation for the approximation $F^k$,
\begin{align*}
&D^k= \sum_{ijlm}\Gamma_{ij}^{lm}\int_\Omega (\frac{F^k_i}{1+\frac{F^k_i}{k}}\frac{F^k_j}{1+\frac{F^k_j}{k}}
-\frac{F^k_l}{1+\frac{F^k_l}{k}}\frac{F^k_m}{1+\frac{F^k_m}{k}})
\ln \frac{F^k_iF^k_j(1+\frac{F^k_l}{k})(1+\frac{F^k_m}{k})}
{(1+\frac{F^k_i}{k})(1+\frac{F^k_j}{k})F^k_lF^k_m}(z)dz.
\end{align*}
Multiply \eqref{Fk-i} by $\ln \frac{F^k_i}{1+\frac{F^k_i}{k}}$, add the equations in $i$, and integrate the resulting equation on $\Omega$. It leads to
\begin{align*}
&\sum_{i= 1}^p\int_{\partial \Omega _i^{-}} \Big( F^k_i\ln F^k_i-k(1+\frac{F^k_i}{k})\ln (1+\frac{F^k_i}{k})\Big) (Z)\mid v_i\cdot n(Z)\mid d\sigma (Z)
+D^k\leq c_b.
\end{align*}
Moreover,
\begin{align*}
k\int _{\partial \Omega _i^{-}} \ln (1+\frac{F^k_i}{k})(Z)\mid v_i\cdot n(Z)\mid d\sigma (Z)&\leq \int _{\partial \Omega _i^{-}}F^k_i(Z)\mid v_i\cdot n(Z)\mid d\sigma (Z)\leq c_b.
\end{align*}
Hence
\begin{align}\label{entropy-flow}
&\sum_{i=1}^p\int_{\partial \Omega _i^-} F^k_i\ln \frac{F^k_i}{1+\frac{F^k_i}{k}}(Z)\mid v_i\cdot n(Z)\mid d\sigma (Z)+D^k\leq c_b.
\end{align}
The uniform entropy dissipation bound holds, since $x\rightarrow x\ln \frac{1+\frac{x}{k}}{x}$ is bounded from above on $] 0,+\infty [ $.
\cqfd
The following lemma replaces an entropy control of $(F^k)_{k\in \N ^*}$, under the condition \eqref{entropy2-main-theorem}.
 %
%
%
%
\begin{lemma}\label{entropy-n0}
Assuming \eqref{entropy2-main-theorem}, it holds that
\begin{align*}
&\sum_{i= 1}^p\int_{z\in \Omega ; F^k_i(z)<k}  F_i^k\ln F_i^k(z)dz+\ln k\sum_{i= 1}^p\int_{z\in \Omega ;F^k_i(z)\geq k}F_i^k(z)dz< c_b,\quad k\in \N ^*,
\end{align*}
where $c_b$ only depends on the given inflow.
\end{lemma}
%
%
\underline{Proof of Lemma \ref{entropy-n0}.}\\
The entropy flow of $(F_i^k)$ is first controlled as follows. It holds that
\begin{align*}
&\int_{\partial \Omega _i^{-}}{F^k_i}\ln (1+\frac{F^k_i}{k})(Z)\mid v_i\cdot n(Z)\mid d\sigma (Z)\\
&\leq \int_{\partial \Omega _i^{-},F_i^k\leq k}{F^k_i}\ln (1+\frac{F^k_i}{k})(Z)\mid v_i\cdot n(Z)\mid d\sigma (Z) \\
&+\int_{\partial \Omega _i^{-},F_i^k\geq k}{F^k_i}\ln (1+\frac{F^k_i}{k})(Z)\mid v_i\cdot n(Z)\mid d\sigma (Z) \\
&\leq \ln 2\int_{\partial \Omega _i^{-}}{F^k_i}(Z)\mid v_i\cdot n(Z)\mid d\sigma (Z)+\int_{\partial \Omega _i^{-},F_i^k
\geq k}{F^k_i}\ln \frac{2F^k_i}{k}(Z)\mid v_i\cdot n(Z)\mid d\sigma (Z)\\
&\leq c_b+\int_{\partial \Omega _i^{-},F_i^k\geq k}{F^k_i}\ln {F^k_i}(Z)\mid v_i\cdot n(Z)\mid d\sigma (Z)-\ln \frac{k}{2}\int_{\partial \Omega _i^{-},F_i^k\geq k}{F^k_i}(Z)\mid v_i\cdot n(Z)\mid d\sigma (Z).
\end{align*}
Together with \eqref{entropy-flow}, this implies that
\begin{align*}
&\sum_{i= 1}^p\int_{\partial \Omega _i^{-},F_i^k\leq k}  F^k_i\ln F^k_i(Z)\mid v_i\cdot n(Z)\mid d\sigma (Z)+\ln \frac{k}{2}\int_{\partial \Omega^{-},F_i^k\geq k}{F^k_i}\mid v_i\cdot n(Z)\mid d\sigma (Z)\leq c_b.
\end{align*}
Set 
\begin{align*}
&e_x:=n_0,\quad \Omega_a=\Omega\cap \{ (x,y); x\leq a\},\quad S_a=\Omega\cap\{(x,y);x=a\},\quad \partial\Omega_a=\partial
 \Omega\cap \bar{\Omega}_a.
\end{align*}  
Multiplying the equation for $F^k_i$ by $\ln\frac{ F^k_i}{1+\frac{F_i^k}{k}}$, $1\leq i\leq p$, summing the resulting equations and integrating over $\Omega _a$, implies that
\begin{align*}
&\sum_{i= 1}^pv_i\cdot n_0\int_{S_a} \Big( F^k_i\ln F^k_i-k(1+\frac{F^k_i}{k})\ln (1+\frac{F^k_i}{k})\Big) (a,y) dy\nonumber \\
&\leq -D_k+\sum_{i= 1}^p\int_{\partial \Omega_a} \Big( F^k_i\ln F^k_i-k(1+\frac{F^k_i}{k})\ln (1+\frac{F^k_i}{k})\Big) (Z) (v_i\cdot n(Z))d\sigma (Z) \nonumber \\
&\leq c_b.
\end{align*}
An integration on $[x_0^-,x_0^+]$ defined in \eqref{df-x0-x0+} implies that
\begin{align*}
&\sum_{i= 1}^pv_i\cdot n_0\int_{\Omega } \Big( F^k_i\ln F^k_i-k(1+\frac{F^k_i}{k})\ln (1+\frac{F^k_i}{k})\Big) (z) dz\leq c_b.
\end{align*}
Moreover,
\begin{align*}
k\int_{\Omega } \ln (1+\frac{F^k_i}{k})(z)dz&\leq \int_{\Omega } F^k_i(z)dz\leq c_b,\quad 1\leq i\leq p,
\end{align*}
and
\begin{align*}
&\int_{\Omega }{F^k_i}\ln (1+\frac{F^k_i}{k})(z)dz\\
&\leq \int_{z\in \Omega ;F_i^k(z)\leq k}{F^k_i}\ln (1+\frac{F^k_i}{k})(z)dz+\int_{z\in \Omega ; F_i^k(z)\geq k}{F^k_i}\ln (1+\frac{F^k_i}{k})(z)dz\\
&\leq \ln 2\int_{\Omega }{F^k_i}(z)dz+\int_{z\in \Omega ,F_i^k(z)
\geq k}{F^k_i}\ln \frac{2F^k_i}{k}(z)dz\\
&\leq c_b+\int_{z\in \partial \Omega ; F_i^k(z)\geq k}{F^k_i}\ln {F^k_i}(z)dz-\ln \frac{k}{2}\int_{z\in \Omega ; F_i^k(z)\geq k}{F^k_i}(z)dz.
\end{align*}
And so,
\begin{align*}
&\sum_{i = 1}^pv_i\cdot n_0\Big( \int_{z\in \Omega,F^k_i(z)<k}  F_i^k\ln F_i^k(z)dz+\ln \frac{k}{2}\int_{z\in \Omega ;F^k_i(z)\geq k}F_i^k(z)dz\Big) < c_b.
\end{align*}
The use of assumption \eqref{entropy2-main-theorem} gives
\begin{align*}
&\int_{z\in \Omega,F^k_i(z)<k}  F_i^k\ln F_i^k(z)dz+\ln \frac{k}{2}\int_{z\in \Omega ;F^k_i(z)\geq k}F_i^k(z)dz< c_b,\quad 1\leq i\leq p,\quad k>2.
\end{align*}
\cqfd
%
%
%
%
%
\section{The passage to the limit in the approximations.}\label{Section4}
\hspace{1cm}\\
\setcounter{equation}{0}
\setcounter{theorem}{0}
This section contains the proof of Theorem 1.1. The  main part is a proof of strong $L^1$ compactness of $(F^k)_{k\in \N ^*}$, based on two compactness lemmas for integrated collision frequency and gain term.
\\
Recall the exponential multiplier form for the approximations $(F^k)_{k\in \N ^*}$,
\begin{align}\label{exponential-form-approximations}
F^k_{i}(z)&= f^k_{bi}(z_i^+(z))e^{-\int_{0}^{s_i^+(z)} \nu_i^k(z_i^+(z)+sv_i)ds}\nonumber\\
&+\int_{0}^{s_i^+(z)} Q_i^{+k}(z_i^+(z)+sv_i)e^{-\int_s^{s_i^+(z)}\nu_i^k(z_i^+(z)+rv_i)dr}ds ,\quad \text{a.a.  }z\in \Omega ,\quad 1\leq i\leq p,
\end{align}
where $\nu _i^k$ and $Q_i^{+k}$ are defined by
\begin{align*}
&\nu _i^k= \sum _{jlm}\Gamma _{ij}^{lm}\frac{F_j^k}{(1+\frac{F_i^k}{k})(1+\frac{F_j^k}{k})},\quad Q_i^{+k}= \sum _{jlm}\Gamma _{ij}^{lm}\frac{F_l^k}{1+\frac{F_l^k}{k}}\frac{F_m^k}{1+\frac{F_m^k}{k}}\hspace*{0.06in}.
\end{align*}
An $i$-characteristics is a segment of points $[Z-s_i^+(Z)v_i,Z]$, where $Z\in \partial\Omega _i^-$.\\
By the strict convexity of $\Omega $, there are for every $i\in \{ 1,\cdot \cdot \cdot p\} $ two points of $\partial \Omega $, denoted by $\tilde{Z_i}$ and $\bar{Z_i}$ such that 
\begin{align*}
z_i^+(\tilde{Z_i})= z_i^-(\tilde{Z_i})\quad \text{and}\quad z_i^+(\bar{Z_i})= z_i^-(\bar{Z_i}).
\end{align*}
Denote by $\Omega^{k2}_{i\epsilon}$ (resp. $\Omega^{k3}_{i\epsilon}$) the set of points between $\tilde{Z_i}$ (resp. $\bar{Z_i}$) and the $i$-characteristics in $\Omega$ at distance $\epsilon$ from  $\tilde{Z_i}$ (resp. $\bar{Z_i}$).
Such subsets of $\Omega $ are introduced in order that all $i$-characteristics from $Z\in \big( \Omega^{k2}_{i\epsilon}\cup \Omega^{k3}_{i\epsilon}\big) ^c$ are segments of length uniformly bounded from below in terms of $\epsilon $.
%
%
\begin{lemma}\label{lemma-df-Omega-epsilon}
\hspace*{1.in}\\
For $k\in \N ^*$, $i\in \{ 1, ..., p\} $ and $\epsilon >0$, there is a subset $\Omega ^{k}_{i\epsilon }$ of $i$-characteristics of $\Omega $ with measure smaller than $c_b\epsilon $, 
containing $\Omega^{k2}_{i\epsilon}$ or $\Omega^{k3}_{i\epsilon}$ defined above, and
such that for any $z\in \Omega \setminus \Omega ^{k}_{i\epsilon }$,
\begin{align}\label{bounds-on-complementary-of-Omega_iEpsilon}
&F^k_{i}(z)\leq \frac{1}{\epsilon }\exp (\frac{1}{\epsilon }),\quad \int_{-s_i^+(z)}^{s_i^-(z)}\nu_i^k(z+sv_i)ds\leq \frac{1}{\epsilon }.
\end{align}
\end{lemma}
\underline{Proof of Lemma 4.1.}\\
It follows from the exponential form of $F^k_i$ that
\begin{align}\label{bdd-by-outgoing}
&F^k_i(z)\leq F^k_i(z+s_i^-(z)v_i)e^{\int _{-s_i^+(z)}^{s_i^-(z)}\nu _i^k(z+rv_i)dr},\quad z\in \Omega .
\end{align}
The boundedness of the mass flow of $(F^k_i)_{k\in \N ^*}$ across $\partial \Omega _i^-$ is
\begin{align*}
&\int _{\partial \Omega _i^-}\mid v_i\cdot n(Z)\mid F^k_i(Z)d\sigma (Z)\leq c_b.
\end{align*}
Consequently, the measure of the set $\{ Z\in \partial \Omega _i^-; F^k_i(Z)>\frac{1}{\epsilon }\} $ is smaller than $c_b\epsilon $. \\
The boundedness of the mass of $(F^k_j)_{k\in \N ^*, 1\leq j\leq p}$ can be written
\begin{align*}
&\int_\Omega F_j^k(z)dz=\int _{\partial \Omega _i^-}\mid v_i\cdot n(Z)\mid \Big( \int _{-s_i^+(Z)}^0F^k_j(Z+rv_i)dr\Big) d\sigma (Z)\leq c_b.
\end{align*}
Hence the measure of the set 
\begin{align*}
&\{ Z\in \partial \Omega _i^-; \int _{-s_i^+
(Z)}^0F^k_j(Z+rv_i)dr>\frac{p^2\Gamma}{\epsilon }\} ,
\end{align*}
where $\Gamma = \max _{i,j,k,l}\Gamma _{ij}^{lm}$, is smaller than $c_b\epsilon $. 
Hence the measure of the set of
$Z\in \partial \Omega _i^-$ outside of which $ F^k_i(Z)\leq\frac{1}{\epsilon } $ and 
$\int _{-s_i^-(Z)}^0F^k_j(Z+rv_i)dr\leq\frac{p^2\Gamma}{\epsilon }$,
is bounded by $c_b\epsilon$.
Together with \eqref{bdd-by-outgoing}, this implies that the measure of the complement of the set of
$Z\in \partial \Omega _i^-$, such that 
\begin{align*}
&F_i^k(z)\leq \frac{1}{\epsilon}\exp \big( \frac{1}{\epsilon}\big) \quad \text{and}\quad \int _{-s_i^+(z)}^{s_i^-(z)}\nu _i^k(z+rv_i)dr\leq \frac{1}{\epsilon}
\end{align*}
for $z=Z-sv_i,\quad 0\leq s\leq s_i^+(Z)$, is bounded by $2c_b\epsilon$. With it $2c_b\epsilon$ is a bound for the measure of the complement, denoted by $\Omega _{i\epsilon }^{k1}$, of the set of $i$-characteristics in $\Omega$ such that for all points $z$ on the $i$-characteristics, 
\begin{align*}
&F_i^k(z)\leq \frac{1}{\epsilon}\exp \big( \frac{1}{\epsilon}\big) \quad  \text{and} \quad  \int_{-s_i^+(z)}^{s_i^-(z)}\nu _i^k(z+rv_i)dr\leq \frac{1}{\epsilon}\hspace*{0.02in}.
\end{align*}
The sets of points $\Omega^{k2}_{i\epsilon}$  and $\Omega^{k3}_{i\epsilon}$ 
have measure of magnitude $\epsilon$, and are also included in $\Omega^k_{i\epsilon}$, 
\begin{align*}
&\Omega _{i\epsilon }^k= \displaystyle{\cup _{p= 1}^3}\Omega _{i\epsilon }^{kp}.
\end{align*}
This ends the proof of the lemma.
\cqfd
Given $i\in \{ 1, ..., p\} $ and $\epsilon$ in Lemma \ref{lemma-df-Omega-epsilon}, let $\chi^k_{i\epsilon }$ denote the characteristic  function of the complement of $\Omega ^{k}_{i\epsilon }$. 
%
%
The following lemma proves the compactness of the $k$-sequence of integrated collision frequencies. 
\begin{lemma}\label{compactness-integrated-collision-frequency}
\hspace*{1.in}\\
The sequences
\begin{align*}
&\Big( \int _0^{s_i^+(z)} \nu _i^k(z_i^+(z)+sv_i)ds\Big) _{k\in \N ^*},\quad 1\leq i\leq p,
\end{align*}
are strongly compact in $L^1(\Omega )$.
\end{lemma}
\underline{Proof of Lemma \ref{compactness-integrated-collision-frequency}.} \\
Let $1\leq i\leq p$. The uniform bound for the mass of $(F^k)$ proven in Lemma \ref{mass-entropy-dissipation-approximations}, implies that 
\begin{align*}
&\int _\Omega \Big( \int _0^{s_i^+(z)} \nu _i^k(z_i^+(z)+sv_i)ds\Big) dz
\end{align*}
is uniformly bounded with respect to $k$.  By
the Kolmogorov-Riesz theorem (\cite{K}, \cite{R}), the compactness will follow from the translational equi-continuity in $L^1(\Omega)$.
The translational equi-continuity in the $v_i$-direction of $\Big( \int _0^{s_i^+(z)} \nu _i^k(z_i^+(z)+sv_i)ds\Big) _{k\in \N ^*}$ follows from the previous uniform bound on $\int _\Omega \Big( \int _0^{s_i^+(z)} \nu _i^k(z_i^+(z)+sv_i)ds\Big) dz$. 
Let us prove the  translational equi-continuity in the $v_j$-direction of each of its terms, 
\begin{align*}
&\Gamma _{ij}^{lm}\int_ {0}^{s_i^+(z)}\frac{F_j^k}{(1+\frac{F_i^k}{k})(1+\frac{F_j^k}{k})}(z_i^+(z)+sv_i)ds.
\end{align*}
It follows from the weak $L^1$- compactness of $(F^k)_{k\in \N ^*}$ that 
\begin{align*}
&\int _\Omega \Big( \int_ {0}^{s_i^+(z)}\Big( (1-\chi _{j\epsilon }^k)\frac{F_j^k}{(1+\frac{F_i^k}{k})(1+\frac{F_j^k}{k})}(z_i^+(z)+sv_i)ds\Big) dz
\end{align*}
can be made arbitrarily small for $\epsilon $ small enough. Consider the remaining term in which $\chi ^k_{j\epsilon}F^k_j$ is bounded by $\frac{1}{\epsilon }\exp (\frac{1}{\epsilon })$.
Noticing that the translational difference  of 
$\frac{F_j^k}{k}$  tends to zero, when $k$ tends to infinity, 
there remains to study the translational difference in the $v_j$-direction of 
\begin{align*}
&\int_ {0}^{s_i^+(z)}\Big( \chi _{j\epsilon }^kF_j^k\Big) (z_i^+(z)+sv_i)ds.
\end{align*}
Write $F_j^k(z_i^+(z)+sv_i)$ in exponential multiplier form,
\begin{align*}
&\int_ {0}^{s_i^+(z)}\big( \chi _{j\epsilon }^kF_j^k\big) (z_i^+(z)+sv_i)ds= A^k_{i,j}(z)+B^k_{i,j}(z),
\end{align*}
where
\begin{align*}
&A^k_{i,j}(z)= \int_ {0}^{s_i^+(z)}
\chi _{j\epsilon }^k(z_i^+(z)+sv_i)f_{bj}^k(z_j^+(z_i^+(z)+sv_i))
e^{-\int_0^{s_j^+(z_i^+(z)+sv_i )}\nu_j^k(z_j^+(z_i^+(z)+sv_i)+\sigma v_j)d\sigma}ds,\\
&B^k_{i,j}(z)= 
\int_ {0}^{s_i^+(z)}\chi _{j\epsilon }^k(z_i^+(z)+sv_i)\int_{0}^{s_j^+(z_i^+(z)+sv_i)}Q_j^{+k}\big( z_j^+(z_i^+(z)+sv_i)+\sigma v_j\big) \\
&\quad \quad \quad \quad \quad e^{-\int_{\sigma }^{s_j^+(z_i^+
(z)+sv_i)} \nu_j^k(z_j^+(z_i^+(z)+sv_i)+\tau v_j)d\tau}d\sigma ds.
\end{align*}
In order to prove the translational equicontinuity of $(A^k_{i,j})$, it is sufficient to prove the translational equicontinuity of 
\begin{align*}
&\Big( \int _0^{s_i^+(z)}\chi _{j\epsilon }^k(z_i^+(z)+sv_i)\int_0^{s_j^+(z_i^+(z)+sv_i )}\nu_j^k(z_j^+(z_i^+(z)+sv_i)+\sigma v_j)d\sigma ds\Big) _{k\in \N ^*},
\end{align*}
 by the $L^1_{v_i\cdot n(Z)}(\partial \Omega ^+)$ compactness of $(f_{bj}^k(z_j^+(z_i^+(z)+sv_i)))_{k\in \N ^*}$.
It is so since, by the change of variables \eqref{change-variables}, each of its terms is a linear combination of
\begin{align*}
&\int _{a_{i,j}(z)}\frac{F_l^k}{(1+\frac{F_j^k}{k})(1+\frac{F_l^k}{k})}(Z)dZ,\quad 1\leq l\leq p,
\end{align*}
with
domains $a_{i,j}(z)\subset \Omega $, continuously depending on $z\in \Omega $,  and such that
\begin{align*}
&\mid a_{i,j}(z)\setminus a_{i,j}(z+h)\mid \leq ch,\quad z\in \Omega ,
\end{align*}
uniformly with respect to $z$.\\
The integral where $F^k_l>\Lambda$, tends to zero when $\Lambda\rightarrow\infty$. If $F^k_l>\Lambda$ at  one but not the other of the two terms in the translation difference, then moving the evaluation points closer, by continuity the larger value of $F^k_l$ can be changed  to $\Lambda$. 
And so we can
assume $F_l^k$ bounded at both evaluation points in the translation difference. It follows 
that $(A^k_{i,j})_{k\in \N ^*}$ is translationally equi-continuous.\\ 
\\
$B^k_{i,j}$ is a sum of
\begin{align*}
&\Gamma _{jj^\prime }^{lm}\int_ {0}^{s_i^+(z)}\chi _{j\epsilon }^k(z_i^+(z)+sv_i)\int_{0}^{s_j^+(z_i^+(z)+sv_i)}\frac{F_l^kF_m^k}{(1+\frac{F_l^k}{k})(1+\frac{F_m^k}{k})}\big( z_j^+(z_i^+(z)+sv_i)+\sigma v_j\big) \\
&\quad \quad \quad \quad \quad e^{-\int_\sigma ^{s_j^+(z_i^+
(z)+sv_i)}\nu_j^k(z_j^+(z_i^+(z)+sv_i)+\tau v_j)d\tau}d\sigma ds
\end{align*}
terms. Consider each  one of these terms and split it into $\Gamma _{jj^\prime }^{lm}(C^k_1+C^k_2+C^k_3)$, where, for real numbers $J_1$ and $J_2$ to be fixed later,
\begin{align*}
&C^k_1(z)= 
\int_ {0}^{s_i^+(z)}\chi _{j\epsilon }^k(z_i^+(z)+sv_i)\int_{0}^{s_j^+(z_i^+(z)+sv_i)}\frac{F_l^kF_m^k}{(1+\frac{F_l^k}{k})(1+\frac{F_m^k}{k})}\big( z_j^+(z_i^+(z)+sv_i)+\sigma v_j\big) \\
&\quad \quad \quad \quad \quad 1_{\frac{F_l^kF_m^k}{(1+\frac{F_l^k}{k})(1+\frac{F_m^k}{k})}\big( z_j^+(z_i^+(z)+sv_i)+\sigma v_j\big) >J_1\frac{F_j^kF_{j^\prime }^k}{(1+\frac{F_j^k}{k})(1+\frac{F_{j^\prime }^k}{k})}\big( z_j^+(z_i^+(z)+sv_i)+\sigma v_j\big) }\\
&\quad \quad \quad \quad \quad e^{-\int_\sigma^{s_j^+(z_i^+
(z)+sv_i)} \nu_j^k(z_j^+(z_i^+(z)+sv_i)+\tau v_j)d\tau}d\sigma ds,\\
&\hspace*{1.in}\\
&C^k_2(z)= 
\int_ {0}^{s_i^+(z)}\chi _{j\epsilon }^k(z_i^+(z)+sv_i)\int_{0}^{s_j^+(z_i^+(z)+sv_i)}\frac{F_l^kF_m^k}{(1+\frac{F_l^k}{k})(1+\frac{F_m^k}{k})}\big( z_j^+(z_i^+(z)+sv_i)+\sigma v_j\big) \\
&\quad \quad \quad \quad \quad 1_{\frac{F_l^kF_m^k}{(1+\frac{F_l^k}{k})(1+\frac{F_m^k}{k})}\big( z_j^+(z_i^+(z)+sv_i)+\sigma v_j\big) <J_1\frac{F_j^kF_{j^\prime }^k}{(1+\frac{F_j^k}{k})(1+\frac{F_{j^\prime }^k}{k})}\big( z_j^+(z_i^+(z)+sv_i)+\sigma v_j\big) , F^k_{j^\prime }\big( z_j^+(z_i^+(z)+sv_i)+\sigma v_j\big) >J_2 }\\
&\quad \quad \quad \quad \quad e^{-\int_\sigma^{s_j^+(z_i^+
(z)+sv_i)} \nu_j^k(z_j^+(z_i^+(z)+sv_i)+\tau v_j)d\tau}d\sigma ds,
\end{align*}
\begin{align*}
&C^k_3(z)=
\int_ {0}^{s_i^+(z)}\chi _{j\epsilon }^k(z_i^+(z)+sv_i)\int_{0}^{s_j(z_i^+(z)+sv_i)}\frac{F_l^kF_m^k}{(1+\frac{F_l^k}{k})(1+\frac{F_m^k}{k})}\big( z_j^+(z_i^+(z)+sv_i)+\sigma v_j\big) \\
&\quad \quad \quad \quad \quad 1_{\frac{F_l^kF_m^k}{(1+\frac{F_l^k}{k})(1+\frac{F_m^k}{k})}\big( z_j^+(z_i^+(z)+sv_i)+\sigma v_j\big) <J_1\frac{F_j^kF_{j^\prime }^k}{(1+\frac{F_j^k}{k})(1+\frac{F_{j^\prime}^k}{k})}\big( z_j^+(z_i^+(z)+sv_i)+\sigma v_j\big) , F^k_{j^\prime }\big( z_j^+(z_i^+(z)+sv_i)+\sigma v_j\big) <J_2 }\\
&\quad \quad \quad \quad \quad e^{-\int_\sigma^{s_j^+(z_i^+
(z)+sv_i)} \nu_j^k(z_j^+(z_i^-(z)+sv_i)+\tau v_j)d\tau}d\sigma ds.
\end{align*}
Using the uniform boundedness of the entropy production terms $(D^k)_{k\in \N ^*}$ and choosing $J_1$ large enough, $C^k_1$ can be made arbitrarily small, uniformly with respect to $k$. For such a $J_1$, notice that
\begin{align*}
C_2^k\leq &\frac{J_1e^{\frac{1}{\epsilon }}}{\epsilon }
\int_ {0}^{s_i^+(z)}\chi _{j\epsilon }^k(z_i^+(z)+sv_i)\int_{0}^{s_j^+(z_i^+(z)+sv_i)}\frac{F^k_{j'}}{1+\frac{F^k_{j^\prime }}{k}}\big( z_j^+(z_i^+(z)+sv_i)+\sigma v_j\big) \\
& 1_{\frac{F_l^kF_m^k}{(1+\frac{F_l^k}{k})(1+\frac{F_m^k}{k})}\big( z_j^+\big( z_j^+(z_i^+(z)+sv_i)+\sigma v_j\big) <J_1\frac{F_j^kF_{j'}^k}{(1+\frac{F_j^k}{k})(1+\frac{F_{j^\prime }^k}{k})}\big( z_j^+(z_i^+(z)+sv_i)+\sigma v_j\big) , F^k_{j^\prime }\big( z_j^+(z_i^+(z)+sv_i)+\sigma v_j\big) >J_2 }\\
&\hspace*{1.8in} e^{-\int_\sigma^{s_j^+(z_i^+
(z)+sv_i)} \Gamma^{lm}_{jj^\prime }\frac{F^k_{j^\prime }}{1+\frac{F^k_{j^\prime }}{k}}(z_j^+(z_i^+(z)+sv_i)+\tau v_j)d\tau}d\sigma ds.
\end{align*}
By the continuity of $F^k$,  the integral with respect to $\sigma$ is a sum of integrals over disjoint intervals where $F^k_{j'}>J_2$.
 The total integral is bounded by an integral of $F^k_{j'}$ over a set where $F^k_{j'}>J_2$. Using the entropy control, $\int_{F^k_{j'}>J_2} F_{j'}^k \rightarrow 0$ when $J_2\rightarrow \infty$. 
And so  choosing $J_2$ large enough, $C^k_2$ can be made arbitrarily small, uniformly with respect to $k$.\\
%
It follows from the boundedness of $(F^k_lF^k_m)$ on its domain of integration in $C^k_3$, that the closing argument in  the proof of translational equi-continuity for $(A^k_{i,j})$ above, can be used to conclude that  $(C^k_3)_{k\in \N ^*}$ 
and with it $(B^k_{i,j})$ are translationally equi-continuous. This ends the proof of the lemma. \cqfd
%
%
%
For any $i\in \{ 1,\cdot \cdot \cdot ,p\}$, the following lemma proves the compactness of  the integrated gain terms
times $\chi ^k_{i\epsilon }$ in the exponential multiplier form of the $(F^k_i)$-sequence, again by the Kolmogorov-Riesz theorem.
\begin{lemma}\label{compactness-integrated-gain-term}
\hspace*{1.in}\\
Take $\epsilon >0$. The sequences 
\begin{align}\label{integrated-gain-term-times-chi}
&\Big( \chi ^k_{i\epsilon }(z)\int_{0}^{s_i^+(z)}
\frac{F_l^kF_m^k}{(1+\frac{F_l^k}{k})(1+\frac{F_m^k}{k})}(z_i^+(z)+s v_i)e^{-\int _s^{s_i^+(z)}\nu _i^k(z_i^+(z)+rv_i)dr}ds\Big) _{k\in \N ^*},\quad i=1,...,p,
\end{align}
are strongly compact in $L^1(\Omega )$.
\end{lemma}
\underline{Proof of Lemma \ref{compactness-integrated-gain-term}.}\\
The sequence $(F^k_i)_{k\in \N ^*}$ being uniformly bounded in $L^1$, the same holds for \eqref{integrated-gain-term-times-chi}. For proving its uniform $L^1$ equi-continuity, split the domain of integration in $(z,s)\in \Omega \times [0,s_i^+(z)] $ into the sets where 
\begin{align*}
&\frac{F_l^kF_m^k}{(1+\frac{F_l^k}{k})(1+\frac{F_m^k}{k})}(z_i^+(z)+sv_i)>J_1\frac{F_i^kF_j^k}{(1+\frac{F_i^k}{k})(1+\frac{F_j^k}{k})}(z_i^+(z)+sv_i),\\
&\big( \text{resp.   }\frac{F_l^kF_m^k}{(1+\frac{F_l^k}{k})(1+\frac{F_m^k}{k})}(z_i^+(z)+sv_i)<J_1\frac{F_i^kF_j^k}{(1+\frac{F_i^k}{k})(1+\frac{F_j^k}{k})}(z_i^+(z)+sv_i)\\
&\hspace*{0.06in} \text{and}\quad F^k_j(z_i^+(z)+sv_i)>J_2\big) ,
\end{align*} 
where the integrals are arbitrarily small for $J_1$ (resp. $J_2$) large enough, and the remaining domain,
\begin{align*}
X:= &\{ (z,s)\in \Omega \times [0,s_i^+(z)] ; \frac{F_l^kF_m^k}{(1+\frac{F_l^k}{k})(1+\frac{F_m^k}{k})}(z_i^+(z)+sv_i)<J_1\frac{F_i^kF_j^k}{(1+\frac{F_i^k}{k})(1+\frac{F_j^k}{k})}(z_i^+(z)+sv_i)\\
&\hspace*{0.1in} \text{and}\hspace*{0.06in} F^k_j(z_i^+(z)+sv_i)<J_2\} ,
\end{align*} 
where $(F^k_lF^k_m)$ is bounded uniformly with respect to $k$. Let us prove the $L^1$ uniform equi-continuity of 
\begin{align*}
&\Big( \chi ^k_{i\epsilon }(z)\int_{0}^{s_i^+(z)}\frac{F_l^kF_m^k}{(1+\frac{F_l^k}{k})(1+\frac{F_m^k}{k})}(z_i^+(z)+s v_i)ds\Big) _{k\in \N ^*}
\end{align*}
on this domain.
We can also restrict to a domain where both $F^k_l(z_i^+(z)+s v_i)$ and $F^k_m(z_i^+(z)+s v_i)$ are bounded, since 
\begin{align*}
&\int _{(z,s)\in X; s\in [0,s_i^+(z)], F^k_l(z_i^+(z)+s v_i)\geq \Lambda }\chi ^k_{i\epsilon }(z)\frac{F_l^kF_m^k}{(1+\frac{F_l^k}{k})(1+\frac{F_m^k}{k})}(z_i^+(z)+s v_i)dsdz\\
&\leq \frac{J_1J_2e^{\frac{1}{\epsilon }}}{\epsilon }\mid \{ (z,s)\in X; s\in [0,s_i^+(z)], F^k_l(z_i^+(z)+s v_i)\geq \Lambda \}\mid 
,
\end{align*}
%
\hspace*{1.in}\\
and the measure of the set where $F^k_l>\Lambda $ tends to zero when $\Lambda \rightarrow +\infty$. And so, we have reduced the problem to proving the $L^1$ uniform equi-continuity of 
\begin{align*}
&\Big( \int _0^{s_i^+(z)}1_{F^k_l(z_i^+(z)+s v_i)<\Lambda }\frac{F^k_l}{1+\frac{F_l^k}{k}}(z_i^+(z)+s v_i)ds\Big) _{k\in \N ^*},
\end{align*}
which follows from the proof of Lemma \ref{compactness-integrated-collision-frequency}. \cqfd
%
%
\begin{lemma}\label{strongL1cv}
\hspace*{1.in}\\
Up to a subsequence $(F^k)_{k\in \N ^*}$ strongly converges in $L^1(\Omega)$. Its limit has finite entropy.
\end{lemma}
%
%
\underline{Proof of Lemma \ref{strongL1cv}.}\\
Let $F$ be a weak $L^1$ limit of a subsequence of $(F^k)_{k\in \N ^*}$. For every $\epsilon >0$, the sequence $(\chi_{i\epsilon}^kF_i^k)_{k\in \N^*}$ is compact in $L^1(\Omega)$ by Lemmas \ref{compactness-integrated-collision-frequency}-\ref{compactness-integrated-gain-term}. 
For a converging subsequence of $(\chi^k_{i\epsilon}F^k)_{k\in \N ^*}$, the limit depends on $\epsilon$.  
Choose a decreasing sequence $(\epsilon_q)$
with $\displaystyle{\lim_{q\rightarrow\infty}}\epsilon_{q}=0$, and a
diagonal subsequence in $k$  with $\chi^k_{i{\epsilon}_q}$ 
converging in $k$ for all $q$,  and increasing with $q$. 
Split $F^k-F$ into 
\begin{align*}
&\chi ^k_{i\epsilon_q}(F^k_i-F_i)+(1-\chi ^k_{i\epsilon_q})F^k_i-(1-\chi ^k_{i\epsilon_q})F_i,\quad 1\leq i\leq p.
\end{align*}
Using that $\int _{\Omega ^k_{i\epsilon_q}}F^k_i$ and $\int _{\Omega ^k_{i\epsilon_q}}F_i$ are arbitrarily small for $\epsilon_q$ small enough, leads to the convergence of $F^k$ to $F$ in $L^1(\Omega )$. Let us prove that $F$ is of finite entropy. It follows from Lemma \ref{entropy-n0} that
\begin{align*}
&\int _{F_i^k(z)\leq k}F_i^k\ln F_i^k(z)dz\leq c_b,\quad k\in \N ^*,\quad 1\leq i\leq p.
\end{align*}
Let $i\in \{ 1,\cdot \cdot \cdot ,p\} $ and $\epsilon >0$ be given. Using Egoroff's theorem, there is a subset $A_\epsilon $ of $\Omega $ such that $(F^k_{i/ A_\epsilon })_{k\in \N ^*}$ uniformly converges to $F_{i/A_\epsilon }$. $F_i$ being continuous is uniformly bounded by some constant $c_\epsilon $ on $A_\epsilon $. For some $k_0\geq 2c_\epsilon $ ,
\begin{align*}
&F_i^k(z)\leq 2c_\epsilon ,\quad k\geq k_0,\quad z\in A_\epsilon .
\end{align*}
Consequently,
\begin{align*}
&\int _{A_\epsilon }F_i^k\ln F_i^k(z)dz\leq c_b,\quad k\geq k_0.
\end{align*}
And so,
\begin{align*}
&\int _{A_\epsilon }F_i\ln F_i(z)dz\leq c_b.
\end{align*}
This implies the boundedness of the entropy of $F$.\\
\cqfd
%
 %
%
%
%
\begin{lemma}\label{renormalized-sol}
\hspace*{1.in}\\
Under the assumptions of Theorem \ref{main-result}, $F$ is a nonnegative renormalized solution of the discrete velocity coplanar Boltzmann boundary value problem \eqref{broad-general-a}-\eqref{broad-general-b}.
\end{lemma}
%
%
\underline{Proof of Lemma \ref{renormalized-sol}.}\\
Start from a renormalized formulation for $\chi ^k_{i\epsilon }F^k_i$,
\begin{align}\label{renorm-1}
&-\int _{\partial\Omega^-}\varphi_i\chi^k_{i\epsilon}\ln \big( 1+F^k_i\big) (Z)v_i\cdot n(Z)d\sigma (Z)- \int_{\partial\Omega^+}\varphi _i\chi^k_{i\epsilon}\ln \big( 1+f^k_{bi}\big) (Z)v_i\cdot n(Z)d\sigma (Z)\nonumber\\
&-\int_{\Omega}\chi^k_{i\epsilon}\ln\big( 1+F^k_i\big)v_i\cdot \nabla \varphi _i(z)dz\nonumber \\
&= \int _{\Omega}\frac{\varphi _i\chi^k_{i\epsilon}}{1+F^k_i}
\sum _{j,l,m=1}^p\Gamma _{ij}^{lm}\Big( \frac{F^k_l}{1+\frac{F^k_l}{k}}\frac{F^k_m}{1+\frac{F^k_m}{k}}-\frac{F^k_i}{1+\frac{F^k_i}{k}}\frac{F^k_j}{1+\frac{F^k_j}{k}}\Big) dz,
\end{align}
for test functions $\varphi\in (C^1(\Omega))^p$.
Use the strong $L^1$ convergence given by Lemma \ref{strongL1cv} for the sequence $(F^{k})_{k\in \N ^*}$, to pass to the limit in the left hand side of \eqref{renorm-1} when $k\rightarrow +\infty $. This gives in the limit for the left hand side
\begin{align*}
&-\int _{\partial \Omega^-}\varphi_i\ln \big( 1+F_i\big) (Z)v_i\cdot n(Z)d\sigma (Z)- \int_{\partial\Omega^+}\varphi _i\ln \big( 1+f_{bi}\big) (Z)v_i\cdot n(Z)d\sigma (Z)\\
&-\int_{\Omega}\ln \big( 1+F_i\big) v_i\nabla \varphi _i(z)dz.
\end{align*}
For the passage to the limit when $k\rightarrow +\infty $ in the right hand side of \eqref{renorm-1}, given $\eta >0$ there is a subset $A_\eta $ of $\Omega$ with $\lvert A_\eta ^c\rvert <\eta $, such that  up to a subsequence, $(F_{k})$ uniformly converges to $F$ on $A_\eta $ and $F\in L^\infty (A_\eta )$. Passing to the limit when $k\rightarrow +\infty $ on $A_\eta $ is straightforward. Moreover,
\begin{align*}
&\lim _{\eta \rightarrow 0}\int _{A_\eta ^c}  \frac{\varphi_i}{1+F_i}Q^-_i(F)(z)dz= 0\quad \text{and}\quad \lim _{\eta \rightarrow 0}\int _{A_\eta ^c}\varphi _i\chi_{i\epsilon }^kF_i^k\nu _i^k(z)dz= 0,
\end{align*}
uniformly with respect to $k$, since
\begin{align*}
&\frac{F_i}{1+F_i}\leq 1,\quad \frac{F^k_i}{(1+F^k_i)(1+\frac{F^k_i}{k})(1+\frac{F^k_j}{k})}\leq 1,\quad \text{and}\quad \lim_{\eta\rightarrow 0}\int_{A^c_\eta}
F^{k}_j =0,
\end{align*}
uniformly with respect to $k$. The passage to the limit in the loss term follows.\\
The passage to the limit in the gain term can be done as follows. The uniform boundedness of the entropy production term of $(F^k)$ given
by \eqref{entropy-flow} in Lemma \ref{mass-entropy-dissipation-approximations}, implies that for any $\gamma >1$,
\begin{align*}
\int _{A_\eta ^c}
 |\varphi _i|\frac{\chi^{k}_i}{1+F^{k}_i}
 \sum _{j,l,m=1}^p\Gamma _{ij}^{lm}\frac{F^{k}_l}{1+\frac{F^{k}_l}{k}}\frac{F^{k}_m}{1+\frac{F^{k}_m}{k}}(z)dz
&\leq \frac{c}{\ln \gamma }+c\gamma \int _{A_\eta ^c}F_i^k\nu _i^k(z)dz.
\end{align*}
Take first $\gamma $ large, then $\eta$ small. 
It follows that the right hand side of \eqref{renorm-1} converges to
\begin{align*}
&\int _\Omega \varphi _i\frac{Q_i^+(F)}{1+F_i}(z)dz-\int _\Omega \varphi _i\frac{Q_i^-(F)}{1+F_i}(z)dz,
\end{align*}
when ${k}\rightarrow +\infty $. Consequently, $F$ satisfies \eqref{broad-general-a}-\eqref{broad-general-b} in renormalized form.
\cqfd
\[\]
\begin{remark}
Strong $L^1$ compactness and convergence to a renormalized solution of the discrete velocity coplanar Boltzmann boundary value problem \eqref{broad-general-a}-\eqref{broad-general-b}, as obtained in Section \ref{Section4}, would also hold without Assumption \eqref{entropy2-main-theorem}, for a sequence of approximations $(F^k)_{k\in \N }$ weakly compact in $L^1$. This will be the frame of a following paper.
\end{remark}
\begin{remark}
Some of the techniques of this paper are used in an ongoing study on the evolutionary Boltzmann equation with coplanar discrete velocities. 
\end{remark}


\begin{thebibliography}{99}
\bibitem
{AN} L. Arkeryd, A. Nouri, {\it On the stationary Povzner equation in $\R^n$},  J. Math. Kyoto Univ. 39 (1) (1999), 115-153.
%
%
\bibitem
{AN2} L. Arkeryd, A. Nouri, {\it Stationary solutions to the two-dimensional Broadwell model}, arXiv 2019, hal-02520758v1.
%
\bibitem
{B} A. Bobylev, {\it Exact solutions of discrete kinetic models and stationary problems for the plane Broadwell model}, Math. Methods Appl. Sci. (4) 19 (1996), 825-845.
%
%
\bibitem
{BPS} A. Bobylev, A. Palczewski, J. Schneider, {\it A consistency result for a discrete-velocity model of the Boltzmann equation}, SIAM J. Numer. Anal. 34 (5) (1997), 1865-1883.
%
%
\bibitem{C} C. Cercignani, Sur des  crit\`eres d'existence globale en th\'eorie cin\'etique discr\`ete, C. R. Acad. Sc. Paris, 301 (1985), 89-92.
%
\bibitem{CIS} C. Cercignani, R. Illner, M. Shinbrot, {\it A boundary value problem for the 2-dimensional Broadwell model}, Commun. Math. Phys. 114 (1988), 687-698.
%
\bibitem{DPL} R. J. DiPerna,  P. L. Lions, {\it On the Cauchy problem for Boltzmann equations: Global existence and weak stability}, Ann. of Math. 130 (1989), 321-366.
%
%
\bibitem{IP} R. Illner, T. Platkowski, {\it Discrete velocity models of the Boltzmann equation: survey on the mathematical aspects of the theory}, SIAM Rev. 30 (1988), 213-255.
%
\bibitem
{Ily} O. V. Ilyin, {\it Symmetries, the current function, and exact solutions for Broadwell's two-dimensional stationary kinetic model}, Teoret. Mat. Fiz. 179 (2014), 350-359.
%
%
\bibitem{K}
A. N. Kolmogorov, {\it \"Uber Kompaktheit der Funktionenmengen bei der Konvergenz im Mittel},
Nachr. Akad. Wiss. G\"ottingen Math.-Phys. KI. II 9 (1931), 60-63.
%
%
\bibitem{R}
M. Riesz, {\it Sur les ensembles compacts de fonctions sommables}, Acta Sci. Math. (Szeged) 6 (1933), 136-142.

\end{thebibliography}
\end{document}